\documentclass[%
 reprint,
superscriptaddress,
 amsmath,amssymb,
 aps,
]{revtex4-2}

\usepackage{graphicx}
\usepackage{dcolumn}
\usepackage{bm}

\begin{document}

\preprint{APS/123-QED}

\title{Non-equilibrium brain dynamics as a signature of consciousness}

\author{Yonatan Sanz Perl}
 \affiliation{Universidad de San Andrés, Buenos Aires, Argentina}%
 \affiliation{Physics Department, University of Buenos Aires, and Buenos Aires Physics Institute, Argentina}%
 \author{Hernán Bocaccio}%
\affiliation{Physics Department, University of Buenos Aires, and Buenos Aires Physics Institute, Argentina}%
\author{Ignacio Pérez-Ipiña}%
\affiliation{Physics Department, University of Buenos Aires, and Buenos Aires Physics Institute, Argentina}%
\author{Steven Laureys}%
\affiliation{Coma Science Group, GIGA Consciousness, University of Liège, Liège,Belgium}%
  \author{Helmut Laufs}%
    \affiliation{Department of Neurology, Christian-Albrechts-University Kiel, Germany}
 \author{Morten Kringelbach}%
  \affiliation{Department of Psychiatry, University of Oxford, United Kingdom}%
  \author{Gustavo Deco}%
 \affiliation{Center for Brain and Cognition, Computational Neuroscience Group, Universitat Pompeu Fabra, Spain}%
 \author{Enzo Tagliazucchi}%
\affiliation{Physics Department, University of Buenos Aires, and Buenos Aires Physics Institute, Argentina}%

\date{\today}

\begin{abstract}
The cognitive functions of human and non-human primates rely on the dynamic interplay of distributed neural assemblies. As such, it seems unlikely that cognition can be supported by macroscopic brain dynamics at the proximity of thermodynamic equilibrium. We confirmed this hypothesis by investigating electrocorticography data from non-human primates undergoing different states of unconsciousness (sleep, and anesthesia with propofol, ketamine, and ketamine plus medetomidine), and funcional magnetic resonance imaginga data from humans, both during deep sleep and under propofol anesthesia. Systematically, all states of reduced consciousness unfolded at higher proximity to equilibrium dynamics than conscious wakefulness, as demonstrated by entropy production and the curl of probability flux in phase space. Our results establish non-equilibrium macroscopic brain dynamics as a robust signature of consciousness, opening the way for the characterization of cognition and awareness using tools from statistical mechanics.

\end{abstract}


\maketitle


\emph{Introduction -}  One of the defining features of living matter is the scale-dependent violation of thermodynamic equilibrium \cite{gnesotto2018broken}. While inanimate matter frequently displays non-equilibrium dynamics, these result from externally applied fields. In contrast, departures from equilibrium in living matter can also arise due to endogenous causes, such as complex chains of enzymatic reactions\cite{fang2019nonequilibrium} driving mesoscopic mechanical processes\cite{battle2016broken}. Another characteristic of non-equilibrium dynamics in living matter is their dependence with the spatial and temporal grain; for instance, cells and larger biological structures might appear to be at equilibrium, while being sustained by non-equilibrium processes at the molecular scale \cite{egolf2000equilibrium}. 

Systems at thermodynamic equilibrium obey detailed balance, meaning that there are no net probability fluxes in the configuration space of the system. Detailed balance leads toward reversible dynamics with null entropy production rate. The vector field representing the probability flux in configuration space becomes purely rotational for steady non-equilibrium systems, the deviation from equilibrium can be quantified using the curl of this field \cite{wang2015landscape}, and entropy production can be estimated from the probability of transitioning between states \cite{roldan2010estimating}. This notion of equilibrium exists only in reference to a certain configuration space, which might reflect, in turn, a particular choice of spatio-temporal grain. 

Models of non-equilibrium brain dynamics have been extensively studied in the literature (including the development of a field theory of stochastic neural activity \cite{buice2007field}) in the contexts of associative learning (i.e. Hopfield networks) and decision making, among others \cite{yan2013nonequilibrium}. The phase space of excitable systems includes the important particular case of scale-free critical behaviour \cite{de2010unstable}. In general, it is agreed that neural dynamics (as many other biological phenomena) are an intrinsically non-equilibrium phenomenon; however, this is less clear for macroscopic brain dynamics \cite{esposito2012stochastic}. A recently proposed framework to quantify entropy production from neural data acquired at a macroscopic scale (functional magnetic resonance imaging [fMRI] recordings) showed that dynamics within a reduced 2D configuration space does not generally obey detailed balance, and that the extent of its violation (and thus the degree of entropy production) is task-dependent \cite{lynn2020non}. 

In spite of these advances, the relationship between conscious states and non-equilibrium dynamics remains to be elucidated. It seems unlikely that the dynamic nature of consciousness and cognition can be sustained by a system without strong deviations from thermodynamic equilibrium at some spatial and temporal scales. Thus, brain states associated with unconsciousness could obey detailed balance at the large scale, while simultaneously being supported by non-equilibrium homeostatic processes at the cellular scale. We investigated electrocorticography (ECoG) recordings acquired during different states of consciousness. Compared to fMRI, this modality is better suited for the estimation of transition probabilities in configuration space, given that higher temporal resolution results in better sampling of transition probabilities. We also investigated fMRI recordings by fitting semi-empirical whole-brain models to data from states of reduced consciousness, using them to produce synthetic time series which arbitrarily high temporal resolution. 


\begin{figure}
\includegraphics[scale=0.235]{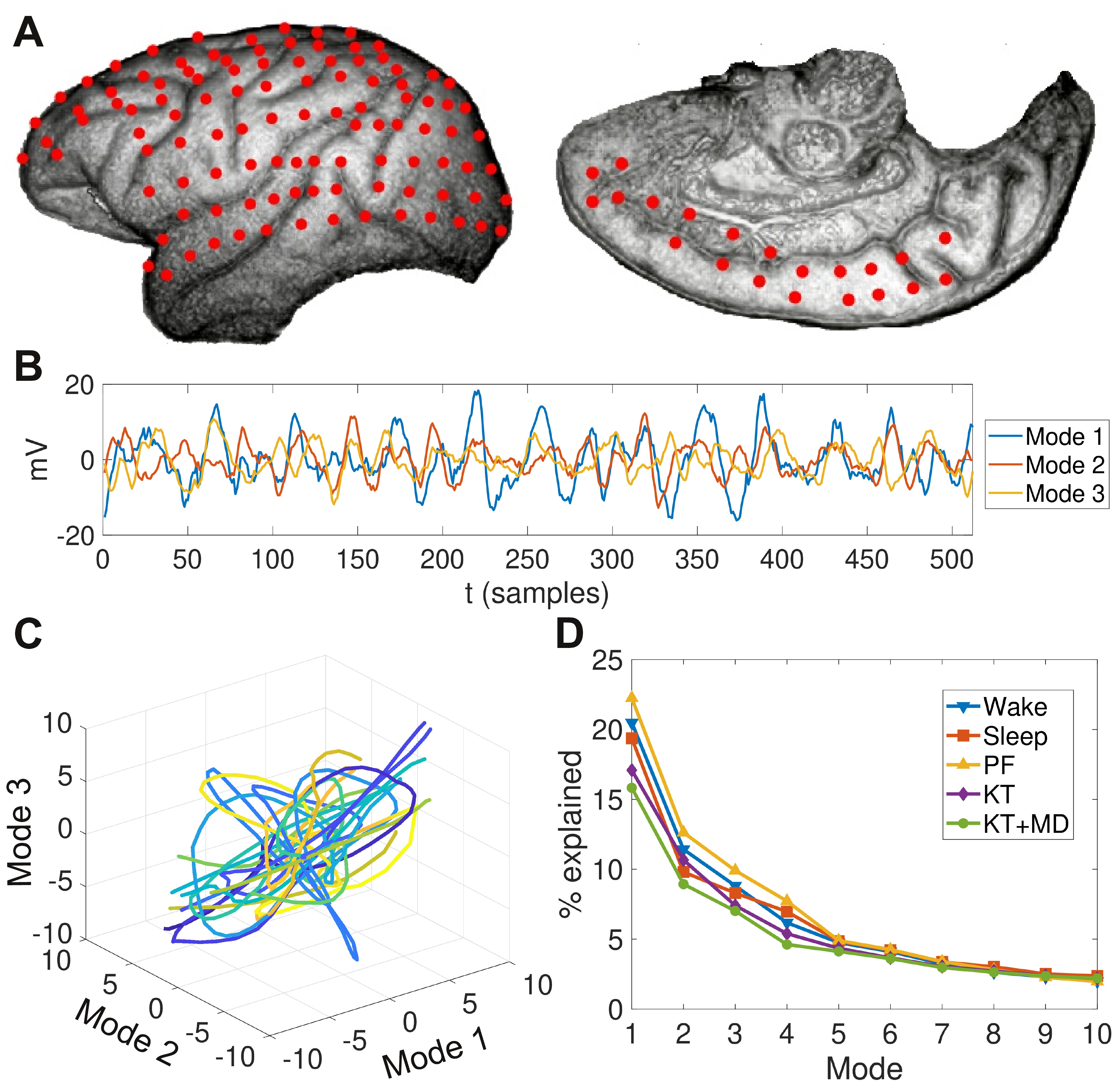}
\caption{\label{method} Principal components of non-human primate ECoG signals. A) Sample ECoG array electrode locations. B) Temporal evolution of the three main modes of ECoG activity for an awake monkey obtained using PCA. C) Time series of panel B projected on the three-dimensional configuration space (the colour gradient represents the temporal evolution of along the trajectory). D) Variance explained as a function of retained principal components, plotted for all states of consciousness.}
\end{figure}

\emph{Consciousness and non-equilibrium dynamics in monkey ECoG data}-
We first considered ECoG data non-human primates (128 channels from electrodes located as shown in Fig. \ref{method}A) during wakefulness, deep sleep (2 animals, 21 sessions with 8 of wakefulness), and under the effects of the following anaesthetic drugs: propofol (PF; 2 animals, 4 sessions of wake, 4 of anaesthesia, 4 of recovery), ketamine  (KT; 2 animals, 4 sessions of wake, 4 of anaesthesia, 4 of recovery), and ketamine plus medetomidine (KF+MD; 4 animals, 11 sessions of wake, 11 sessions of anaesthesia, and 10 sessions of recovery). All sessions were recorded with eyes closed. Time series were notch filtered to remove the power line frequency and its first harmonics (50 Hz, 100 Hz, 150 Hz) and then band-pass filtered between 5 and 500 Hz. Filtered signals were downsampled from 1 kHz to 256 Hz using linear interpolation prior to application of PCA. For more information on this data set visit \url{http://neurotycho.org/}. 

Sample results obtained from linear dimensional reduction are shown in Fig. \ref{method}B as the temporal evolution of the first three modes obtained applying PCA to ECoG of an awake animal. Fig. \ref{method}C shows the projection of these time series on the space defined by the three principal modes, with temporal evolution represented by a colour gradient from blue scales (earlier times) to yellow scales (later times). The percentage of the variance explained as a function of retained principal components is shown in Fig \ref{method}D separately for all states of consciousness.     

\begin{figure}
\includegraphics[scale=0.45]{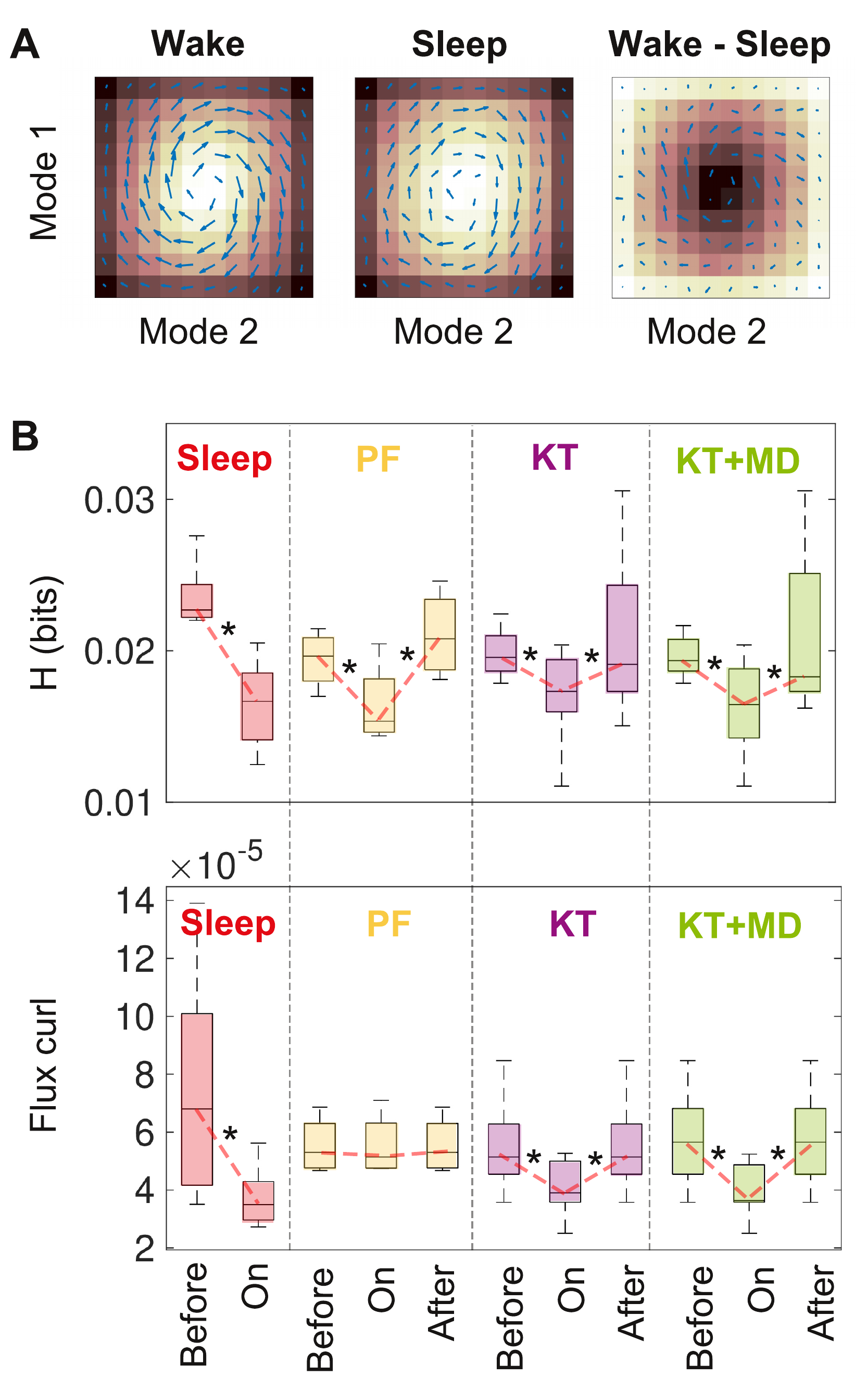}
\caption{\label{monos} Non-equilibrium entropy production is a signature of conscious states. A) 2D configuration space defined by the two main ECoG modes for sleep, wake and their subtraction. B) Up: entropy production per state and condition. Bottom: curl of the probability flux per state and condition.*p$<$0.05, Wilcoxon’s test, Bonferroni corrected for multiple comparisons.}
\end{figure}

We then assessed the violation of detailed balance by dividing the two-dimensional space spanned by the two main modes of ECoG activity into $10 \times 10$ grids from -2 to 2 standard deviations (s.d.) \cite{battle2016broken,lynn2020non}. Figure \ref{monos}A presents the average probability density and flux in configuration space for wake, sleep and their subtraction. 

We then estimated the entropy production as a measure of equilibrium/non-equilibrium behaviour. We associated with each brain state a  probability transition $P_{ij}$ representing the probability of transitioning from state $x_i$ at time $t$ to $x_j$ at time $t+1$, where the state of the system is defined by the grid in the configuration space defined by the two main modes. The arrows in Fig. \ref{monos}A are scaled to the transition probabilities $P_{ij}$ between adjacent states; rotational fluxes are typical of non-equilibrium dynamics and result in net entropy production. We calculated the entropy production as: $S=\sum_{ij}P_{ij}log(\frac{P_{ij}}{P_{ji}})$ \cite{roldan2010estimating,lynn2020non}. Figure \ref{monos}B (up) shows that entropy production is minimal during states of unconsciousness (sleep, KT, PF, KT+MD) compared to wakefulness before and after anaesthesia, i.e. recovery. Figure \ref{monos}B (bottom) presents the curl of the probability flux, significantly larger for all states of unconsciousness vs. controls with the exception of PF.  

\begin{figure}
\includegraphics[scale=0.235]{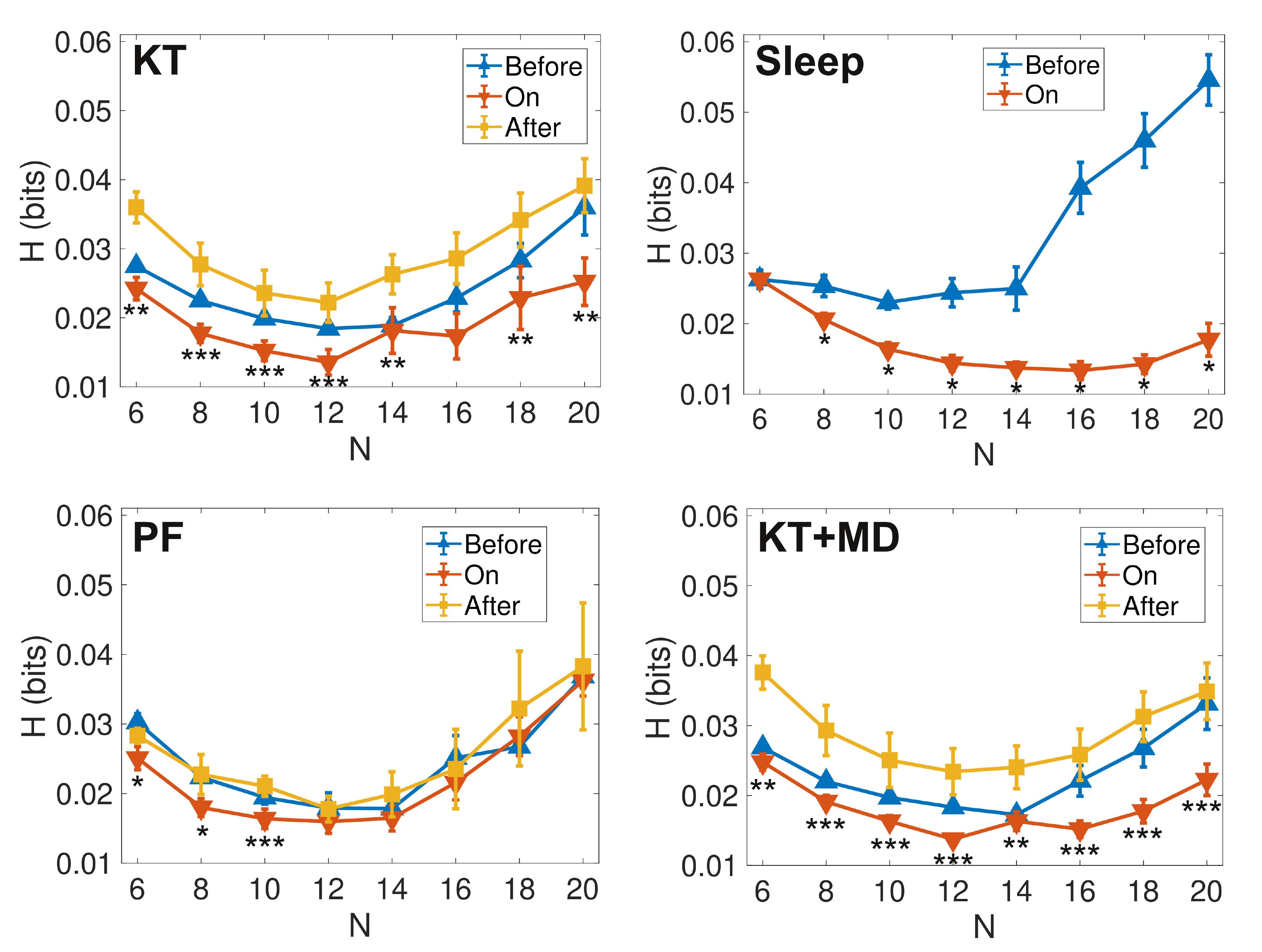}
\caption{\label{grain} Entropy production for different grid sizes. Each panel presents entropy production as a function of the the square root of the number of cells for a given condition (sleep, PF, KT, KT+MD). *p$<$0.05 before vs. on, **p$<$0.05 after vs. on, **p$<$0.05 after and before vs. on, Wilcoxon’s test, Bonferroni corrected for multiple comparisons.}
\end{figure}

\emph{Changes in grid size}- We repeated the procedure for grids with different numbers of cells, from $N=6^2$ to $N=20^2$ cells, and computed entropy production for each size. Figure \ref{grain} shows the results obtained for sleep, KT, PF and KT+MD, as well as for the corresponding control conditions. Entropy production presented a minimum value when phase space for $N=10-12$ cells by and then increased monotonically. Significant differences were observed for most grid sizes.

 \emph{Second order transition probabilities}- The calculation of entropy production by means of probability flux in configuration space assume Markovian dynamics \cite{roldan2010estimating}. This represents an important limitation in our analyses, since higher order temporal dependencies have been suggested by prior reports. To assess the potential impact of this limitation, we included a second order term in the computation of the transition probabilities, i.e. $P_{i,j,k}$ \cite{lynn2020non}. The results preserved all significant differences shown in Fig. \ref{monos}.

\begin{figure}
\includegraphics[scale=0.30]{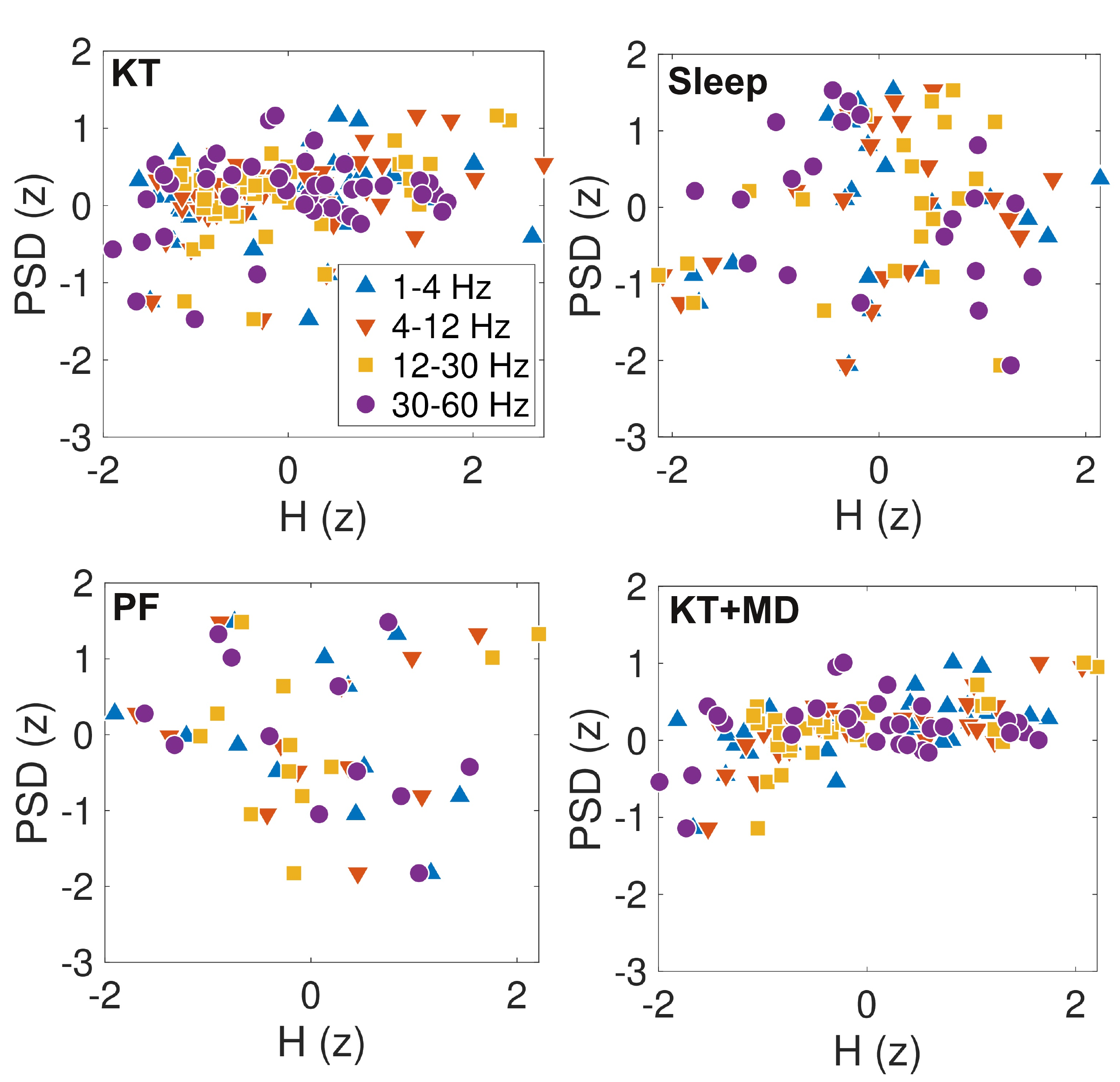}
\caption{\label{pws} Comparison between entropy production and spectral content for frequency bands 1-4 Hz, 4-12 Hz, 12-30 Hz, 30-60 Hz, indicated with different symbol shapes and colours. No significant correlation between these variables were found.}
\end{figure}

 \emph{Entropy production vs. spectral content}- We compared the assessment of non-equilibrium dynamics by means of entropy production with the spectral power in different bands (1-4 Hz, 4-12 Hz, 12-30 Hz, 30-60 Hz), an empirical metric that is frequently used to assess (un)consciousness in electrophysiological signals \cite{koch2016neural}. Figure \ref{pws} presents scatter plots of spectral content in different bands (indicated by symbol type and colour) vs. entropy production for all conditions. We did not observe significant correlations that could be interpreted as redundant information between spectral content and entropy production.

\emph{fMRI data enhancement using whole-brain semi-empirical models}- 
To investigate non-equilibrium dynamics as a signature of consciousness in human fMRI recordings, we first developed a data enhancement technique allowing us to improve the sampling of configuration space. 

For this purpose, we implemented computational models linking together multiple sources of empirical data by means of coupled local dynamics with different qualitative behaviours \cite{cofre2020whole}. Our model incorporates empirical estimates of the large-scale structural connectivity of the brain obtained using diffusion tensor imaging (DTI), a non-invasive methodology to estimate axon bundles from water diffusion anisotropies. We computed observable based on the large-scale functional connectivity of the brain, i.e. the linear correlation between mean all pairs of fMRI time series extracted from 90 pre-defined regions of interest spanning the whole cortical and sub-cortical grey matter \cite{tzourio2002automated}. Finally, we constructed models with heterogeneous local parameters whose variation is constrained by anatomical priors that represent boundaries between brain regions associated with specific functional systems (resting state networks [RSN]) \cite{ipina2020modeling}.

\begin{figure}
\includegraphics[scale=0.5]{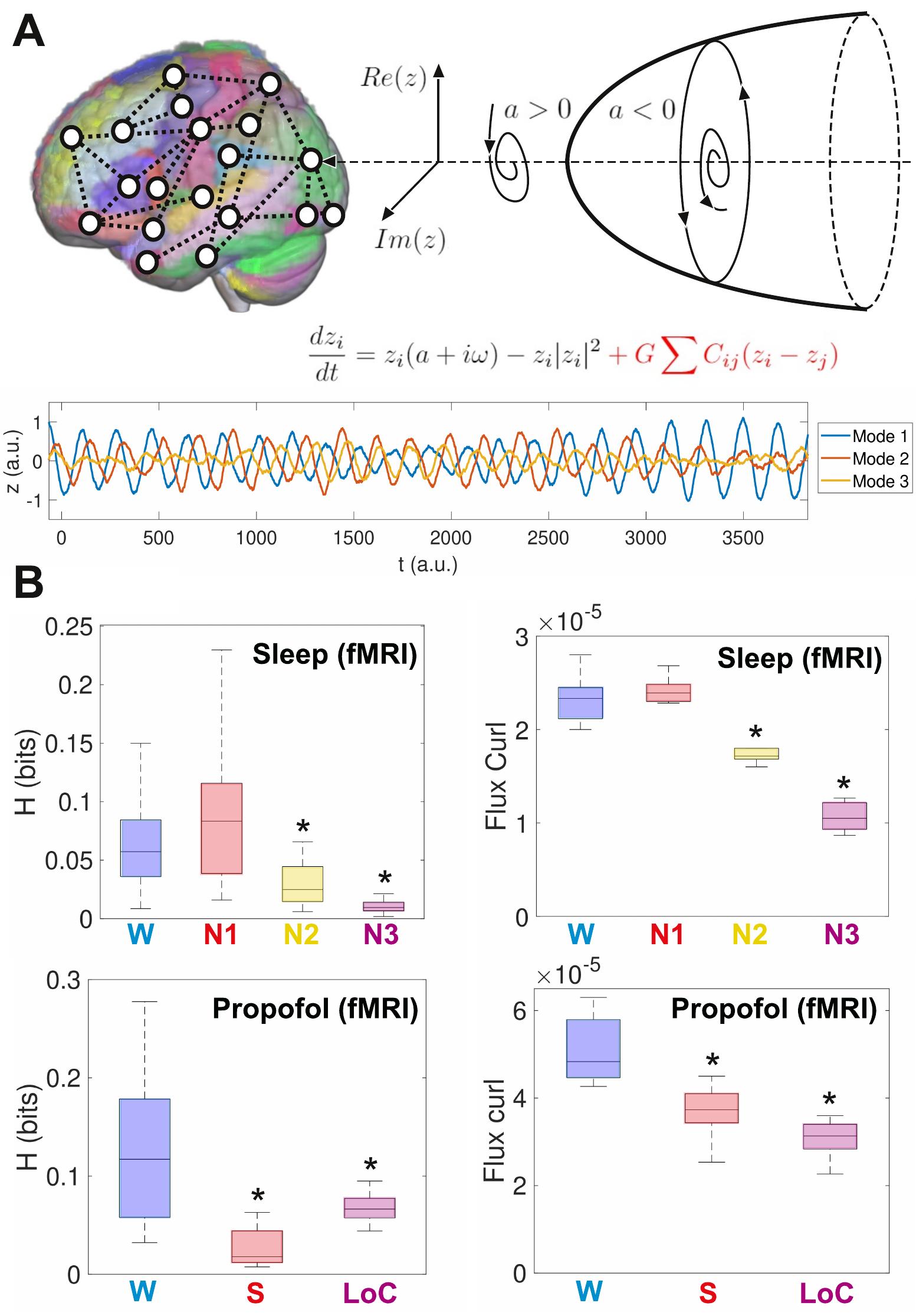}
\caption{\label{augmented} Model-based fMRI data enhancement used to assess the non-equilibrium dynamics in the human brain. A) Up: a whole-brain semi-empirical model presenting dynamical criticality (local Hopf bifurcation) fitted to the empirical functional connectivity of different sleep stages, using structural connectivity (DTI) to couple the local dynamics, and different RSN as priors to constraint the independent variation of local bifurcation parameters. Bottom: the temporal evolution of the three principal modes obtained using PCA over the simulated time series. B) Entropy production (left) and flux curl (right) computed from the model fitted to empirical fMRI data obtained during wakefulness and three progressively deeper stages of human sleep (up), as well as  or wakefulness and two levels of propofol-induced loss of consciousness (bottom).*p$<$0.05, Wilcoxon’s test, Bonferroni corrected for multiple comparisons.}
\end{figure}

We outline this model in Fig.~\ref{augmented}A.  Local dynamics present dynamical criticality when a region undergoes a bifurcation from stable fixed point dynamics (a$<$0) to oscillatory behaviour (a$>$0) according to a Hopf bifurcation; these dynamics are coupled by the structural connectivity network inferred from DTI data, and scaled by a global coupling parameter ($G$). As shown in \cite{ipina2020modeling} this scaling factor $G$ and the local bifurcation parameters  corresponding to each anatomical prior (i.e. $a$ value for each RSN) can be optimized to reproduce the functional connectivity measured during different states of consciousness. 

We applied this optimization procedure to reproduce the empirical observables for wakefulness, for all human non-rapid eye movement (NREM) sleep stages (N1, N2, N3), and for propofol-induced sedation and loss of consciousness. We then used the optimal parameters to enhance the BOLD signal lengths of each subject (n=15) up to 30.000 samples for each region of interest \cite{perl2020data, perl2020generative}, allowing us to estimate entropy production as done with the ECoG data (Fig. \ref{monos}). Figure \ref{augmented}A (bottom) presents the three main modes extracted from the simulated BOLD time series using PCA for parameters corresponding to wakefulness. Figure \ref{augmented}B presents entropy production and curl of probability flux estimated from the time series corresponding to all sleep stages and depths of propofol sedation. We found that N2 and N3 sleep (the deepest stages of NREM sleep) presented dynamics significantly closer to detailed balance than conscious wakefulness; we found the same for propofol sedation and loss of consciousness, but not for early (N1) sleep.

\emph{Conclusions}- The combined assessment of ECoG and fMRI data (assisted by model-based temporal enhancement) demonstrated that macroscopic non-equilibrium dynamics are a signature of consciousness spanning two model organisms measured with two different neural activity recording techniques, both during deep sleep (N2 and N3 sleep, but not N1), and under the effects of several anaesthetic drugs (propofol, ketamine, and ketamine plus medetomidine for monkeys; two doses of propofol for humans).

Most of the leading theories agree that consciousness and cognition emerge from the coordinated and dynamic interplay of distributed brain activity. For example, the global workspace theory posits that consciousness is equivalent to the non-linear percolation of sensory information throughout an anatomical connectivity backbone, allowing flexible access by different modular cognitive processors \cite{mashour2020conscious}. The dynamic core hypothesis by Edelman and Tononi identifies consciousness with an integrated dynamic process capable of achieving an enormous number of configurations \cite{tononi1998consciousness}. Even if these explanations are considered satisfactory from a mechanistic perspective, the problem stands still: how does the neural collective self-organizes to fulfil the dynamic requirements of these theories? Our work shows that non-equilibrium brain dynamics is a general feature of conscious states. In turn, this type of dynamics can be achieved through different mechanisms, such as statistical \cite{tagliazucchi2016large} and dynamic criticality \cite{solovey2015loss}. 

Importantly, we established the independence between non-equilibrium entropy production and spectral features that are frequently used to characterize unconscious states. This observation is relevant since some of the general anaesthetic we investigated (e.g. ketamine \cite{maksimow2006increase}) fail to induce the type of low frequency and complexity oscillations that considered indicative of unconsciousness, suggesting that our metrics are more general. 

In summary, we demonstrated a link between a very general property of the brain as a macroscopic physical system, and the emergence of consciousness and cognition. Future studies should refine our conclusions, attempting to converge towards the relationship between dynamics and computation in neural tissue, one of the most challeing and long-stading problems in the field. 
\\
\\
Authors acknowledge funding from Agencia Nacional De Promocion Cientifica Y Tecnologica (Argentina), grant PICT-2018-03103. The authors acknowledge  the Toyoko 2020 program for granting cloud computing services.

\bibliography{apssamp}

\end{document}